\documentclass[conference]{IEEEtran}
\IEEEoverridecommandlockouts
\usepackage{cite}
\usepackage{amsmath,amssymb,amsfonts}
\usepackage{algorithmic}
\usepackage{enumitem}
\usepackage{graphicx}
\usepackage{textcomp}
\usepackage{xcolor}
\usepackage{color,soul}
\usepackage{multirow}
\usepackage{booktabs}
\usepackage{hyperref} 
\def\BibTeX{{\rm B\kern-.05em{\sc i\kern-.025em b}\kern-.08em
    T\kern-.1667em\lower.7ex\hbox{E}\kern-.125emX}}
\begin{document}

\title{
An Agentic RAG and Evaluation Framework for Assurance Case Generation: Industrial Use Case for the EU Cyber Resilience Act Compliance
% *\\
% {\footnotesize \textsuperscript{*}Note: Sub-titles are not captured for https://ieeexplore.ieee.org  and
% \thanks{EU Horizon No. 101120606.}
}

\author{
\IEEEauthorblockN{
Fariz Ikhwantri\IEEEauthorrefmark{4},
Iker Lasa Ojanguren\IEEEauthorrefmark{2},
Dusica Marijan\IEEEauthorrefmark{4},
Maria I. Maslioukova\IEEEauthorrefmark{3},
José Arias Marin\IEEEauthorrefmark{2},
Pavlos Kosmides\IEEEauthorrefmark{3}
}

\IEEEauthorblockA{\IEEEauthorrefmark{4}
Simula Research Laboratory, Kristian August Gate 23, 0164, Oslo, Norway\\
Emails: \{fariz, dusica\}@simula.no
}

\IEEEauthorblockA{\IEEEauthorrefmark{2}
TECNALIA, Astondo Bidea 700, Derio 48160, Biscay, Spain\\
Emails: \{iker.lasa, jose.arias\}@tecnalia.com
}

\IEEEauthorblockA{\IEEEauthorrefmark{3}
Catalink Limited, Kleomenous 2, 1061, Nicosia, Cyprus\\
Emails: \{maria.maslioukova, pkosmidis\}@catalink.eu
}
}

\maketitle

\begin{abstract}
Complying with the EU Cyber Resilience Act (CRA) is a resource-intensive challenge for SMEs due to the complexity of cybersecurity conformity assessment. Yet, it is essential for demonstrating regulatory compliance and ensuring product security and resilience. To address this, we introduce an automated framework for generating Assurance Cases (ACs) using an agentic Retrieval-Augmented Generation grounded in a formal Claim-Argument-Evidence logic. By systematically mapping technical documentation requirements, the framework streamlines the generation of certification evidence. We validate our approach on a case study of Catalink’s PATROLIoT wildfire monitoring system, where the agentic RAG generated 70 ACs with high grounding density ($\approx$4.4 artefacts per AC). The proposed Natural Language Inference (NLI) evaluator achieves 0.88 accuracy, which provides traceability, while expert-validated plausibility (3.06) supports interpretable justifications. For practitioners, this work provides a scalable, interpretable approach for automating mandatory CRA conformity assessments, where it automates document retrieval and AC construction steps that are otherwise performed manually while maintaining transparent decision support.
\end{abstract}


\section{Introduction}

Assurance cases (ACs) are a structured form of reasoning used in safety- and security-critical requirements engineering, typically expressed as tree-structured argumentation (e.g., Claims–Arguments–Evidence)~\cite{rushby2015understanding}. They improve confidence in system requirements by providing evidence of traceability, logical consistency, and completeness~\cite{MANSOUROV201123}. However, constructing high-quality AC is a labour-intensive and expertise-driven process, creating a bottleneck in the development and certification of complex systems for companies~\cite{10.1145/3722571.3727837}.

The introduction of the EU Cyber Resilience Act (CRA) further increases the need for efficient assurance engineering. The CRA establishes mandatory cybersecurity requirements for products with digital elements placed on the EU market. It requires manufacturers to demonstrate compliance through comprehensive technical documentation and supporting evidence. While these requirements improve the security and resilience of digital products, preparing conformity evidence imposes a substantial burden on small and medium-sized enterprises (SMEs), which often lack dedicated compliance expertise and resources. Reducing the effort required to develop transparent and traceable assurance arguments is therefore an important industrial challenge.

Prior study has utilised structured patterns and Large Language Models (LLMs) to automate the synthesis and evaluation of assurance arguments~\cite{sivakumar2024prompting, ODU2025112353, 10771339, 10.1145/3691620.3695296}. While these approaches demonstrate the potential for automation, they are often restricted to the instantiation of predefined templates or suffer from a limited scope that does not account for the heterogeneous and fragmented technical documentation required for industrial certification. Consequently, establishing deep, verifiable traceability between generated claims and the specific local evidence remains a significant challenge.

Furthermore, industrial adoption is hindered by significant practical barriers. Existing frameworks often assume cloud-based access to LLM services, posing confidentiality risks for proprietary documentation during the certification process~\cite{sivakumar2024prompting, chen2025trusta}. On the evaluation side, automated review such as "LLM-as-a-judge" frequently struggles with the semantic nuances of technical reasoning~\cite{yu2025llmsjudgesautomaticreview} or depends on expensive, manually annotated datasets~\cite{ikhwantri2026evaluatingassurancecasestextattributed} that are unavailable for new regulatory domains like the EU CRA. These limitations highlight an urgent need for an end-to-end framework capable of grounded generation and interpretable verification under realistic industrial constraints. Our work addresses these gaps by integrating an agentic RAG approach with a domain-expert plausibility study, focusing on the utility of NLI-based rationales in a regulatory environment where the training data is not readily available.

To address these challenges, we propose an end-to-end framework for automated assurance case generation based on Agentic Retrieval-Augmented Generation (Agentic RAG). The framework constructs assurance cases using the Claim–Argument–Evidence (CAE) structure while grounding each generated argument in organisation-specific technical documentation retrieved from controlled local knowledge sources. Rather than verifying the correctness of the underlying evidence itself, our approach focuses on verifying the logical consistency of the generated assurance text components. Specifically, it assesses whether generated claims are semantically supported by their associated evidence and subclaims. This enables transparent and interpretable assessment of generated assurance cases while remaining compatible with industrial confidentiality requirements.

To evaluate the quality of the generated ACs, we integrate an interpretable NLI-based evaluation~\cite{stacey2022supervising, vladika2025step, ikhwantri2025explainable} framework that measures semantic support between generated claims, subclaims, and evidence. In addition, we complement the automatic evaluation with a domain expert plausibility study to reflect the industrial assurance needs better.

We validate our framework in an industrial case study of Catalink’s PATROLIoT wildfire monitoring system, where we generate ACs covering key requirements aligned with the EU CRA. In total, 70 assurance cases spanning seven CRA requirements are automatically constructed and evaluated. The framework achieves $\approx$4.4 grounded artefacts per AC, while the NLI evaluator reaches 0.88 accuracy and 0.85 F1 on unseen auditor data.
% The remainder presents related work, methodology, evaluation, and conclusions.
% This demonstrates the feasibility of applying grounded, agentic generation and interpretable evaluation in a real industrial compliance scenario under realistic documentation constraints. 

% 

% 

% To ensure rigorous evaluation, we propose a complexity-controlled experimental framework that varies assurance case structure, including the number of components, argumentation depth, and requirement coverage. This approach enables analysis of how well distilled models maintain reasoning quality as structural complexity increases, rather than relying exclusively on manually constructed data comparisons. We also perform evaluations that incorporate automated structural validation, logical consistency checks, and expert assessment to provide a comprehensive perspective on model performance.

\section{Related Work}

\subsection{Assurance Case Generation}

There is increased interest in automating assurance case generation to mitigate the high cost and manual effort of certification. Early approaches utilised safety patterns and reusable templates~\cite{NESIC2021110922, Porter_2023}, which guide argument construction but require significant expert customisation. To increase autonomy, research has shifted toward LLMs using sophisticated prompting strategies to synthesise Goal Structured Notation (GSN) structures from high-level requirements~\cite{sivakumar2024prompting, ODU2025112353}. Recent advancements also explore hybrid formal methods like TrustAs~\cite{chen2025trusta} and the automation of \textit{eliminative argumentation}, which strengthens arguments by identifying potential weaknesses or ``defeaters.'' Frameworks such as \textit{CoDefeater}~\cite{10.1145/3691620.3695296} and AI-supported eliminative systems~\cite{10771339} leverage LLMs to automatically generate candidate defeaters, helping engineers iteratively address vulnerabilities in safety and security claims.

Despite the generative capabilities, a critical industrial bottleneck persists: most approaches rely on internal LLM knowledge rather than grounded product artefacts, yielding arguments that are syntactically valid but semantically unsound or factually misaligned~\cite{Yahya}. Since industrial documentation is typically scattered and heterogeneous, existing frameworks often fail by assuming pre-structured templates. Navigating unstructured repositories under SME constraints requires an agentic approach for iterative, grounded search and refinement to ensure local evidence substantiates every claim.

\subsection{Assurance Case Evaluation}

AC evaluation has evolved from qualitative structural criteria~\cite{chowdhury2020systematic} and graph-based analysis~\cite{ikhwantri2026evaluatingassurancecasestextattributed} toward semantic verification. Early efforts in version comparison~\cite{kobayashi2016comparison} and more recent multi-hop NLI approaches~\cite{ikhwantri2025explainable} focus on verifying whether evidence logically entails a claim, moving beyond syntax to analyse reasoning integrity. Most recently, the "LLM-as-a-judge" paradigm has been proposed to provide automated review against rubrics~\cite{yu2025llmsjudgesautomaticreview}. However, as Yu et al. noted, LLM judges often provide vague assessments due to a lack of access to the external technical context. Our work bridges this gap by extending our previous multi-hop NLI framework~\cite{ikhwantri2025explainable} with Agentic RAG generation, synthetic-to-real evaluation, and an industrial CRA case study.
% Our work bridges this gap by {grounding the evaluation in an agentic RAG framework, utilising fine-grained NLI and token-level heatmaps to align automated judgments with domain expert evaluation. 

\section{Industrial Use Case: Cyber Resilience Act in Fire Detection System}

PATROLIoT is an Internet of Things (IoT) system designed to support firefighters and first responders in monitoring rural areas and detecting potential wildfire events. The system is equipped with environmental sensors that provide situational awareness, as well as a camera that offers visual feedback from the monitored area. Captured images are processed locally using machine learning models to detect the presence of fire or smoke and generate alerts in a user dashboard to support timely response actions.

As an IoT system transmitting data to a backend, PATROLIoT falls under the CRA. Since it does not meet the CRA's Annex III/IV criteria for 'important' or 'critical' products, it requires self-assessment, shifting the full burden of demonstrating compliance with essential requirements onto the SME. In practice, self-assessment does not make the process straightforward for an SME. Without dedicated cybersecurity or regulatory compliance personnel, it is often unclear which existing technical documentation qualifies as relevant evidence, how it should be organised to support each requirement, and how to trace it systematically back to specific CRA articles. Relevant information is typically scattered across design documents, test reports, and internal records. This makes the preparation of assurance arguments a slow, expertise-dependent task and a significant barrier to timely compliance.

\section{Methodology}

\subsection{Agentic RAG: Grounded Assurance Case Generation}

% IKER TODO: introduce the subject for certification (product) JSON description that is injected at runtime.
% Assurance Case generation is cast as an agentic retrieval-augmented generation (RAG) problem in which a regulatory requirement anchors a structured argumentation artefact. In the use case, a Cyber Resilience Act requirement (e.g., CRA-A1.P1.1) is selected at session start. During construction, a ReAct~\cite{yao2023react} agent produces searches based on requirements and product characteristics. These sequential searches are refined accordingly by the model itself as more content is retrieved. Once the base model considers that enough information has been gathered, the agent produces a tree-like JSON structure following a Claim-Argument-Evidence target schema. The produced artefact showcases requirement decomposition capabilities that come together with the generation of claims, arguments, and evidence grounded in the industrial product documentation.

We approach assurance case generation as an RAG problem combined with agentic AI. The agent constructs an assurance case grounded in regulatory requirements with multiple iterations. We use Agentic AI based on the ReAct method~\cite{yao2023react} from LangChain\footnote{https://github.com/langchain-ai/langchain} to support iterative retrieval and query refinement over heterogeneous technical documentation. The agent starts with i) a requirement from the CRA and ii) the product context information: a JSON string of less than 1800 characters that covers a short product description, component type, deployment context, firmware OS, protocols, security features and certification scope. Then, the agent iteratively generates search queries informed by both the requirement and product characteristics. As the agent retrieves additional information, it dynamically refines its subsequent queries to maximise relevance and coverage. When the agent determines that sufficient supporting information has been collected, it synthesises the findings into a hierarchical JSON artefact that adheres to a Claim-Argument-Evidence tree schema.
This artefact reflects the decomposition of the initial requirement and grounds each claim, argument, and evidence item in the context of industrial product documentation.

\noindent\textbf{Agent Instruction.}
%certification context has not been mentioned before.
The agent is steered by a comprehensive system prompt that ties together \textit{the CRA requirement, the product context information, retrieval tools, and the assurance-case output format}. Session-specific fields (the standard, requirement ID and text, and product context) are passed at runtime. Recent conversation and retrieval activity can be appended so that each turn reflects the current session, not only the static instructions. The prompt is split into a few sections: \emph{what} the agent must produce (CAE-based JSON with requirement traceability), \emph{how} it should behave (evidence-grounded, explicit about gaps, no persistence without user confirmation) and \emph{which tools} to use for document search and optional code sources.

% This hybrid design enables the agent to perform broad semantic exploration when the evidence location is uncertain, while allowing for progressively narrower searches when document types, approval statuses, or filename patterns are known. Such behaviour is essential for correlating Cyber Resilience Act (CRA) obligations with heterogeneous artefacts—such as design documents, test reports, and audit records—ingested from industrial workflows.

\noindent\textbf{Agentic Tools.} Our agent architecture follows a hybrid design that integrates both (a) document ingestion and (b) semantic search tool functionalities. 

\textbf{(a) Document ingestion} processes and organises heterogeneous artefacts, such as design documents, test reports, and audit records, from industrial workflows. Documentation is ingested via \textit{Docling}, segmented into 3,000-character chunks with a 500-character overlap, and indexed in ChromaDB using BGE-M3 embeddings. 

\textbf{(b) Semantic search tools} enable broad semantic exploration when the evidence location is unknown and support targeted, filtered retrieval based on exact content match and/or filename patterns. An additional tool collects file names and relevant metadata in advance, allowing the agent to contextualise the documentation and leverage filename semantics. These also help to inform and refine subsequent searches. The agent executes semantic search via L2 distance, utilising: (i) lexical constraints for keyword-anchored retrieval, and (ii) a two-phase filename filter that identifies target documents prior to querying to prevent post-retrieval truncation. These tools ensure high recall across the heterogeneous logs and specifications common in industrial workflows.

\noindent\textbf{Output argumentation schema.}
\label{sec: output-schema}
The agentic RAG produces a hierarchical JSON artefact following a CAE schema. The root \textit{Claim} node links to a CRA ID; sub-claims decompose via \textit{Argument} objects (AND/OR/ARG). Terminal nodes list \textit{Evidence} objects, which capture descriptive content, epistemic status (e.g., type, confidence scores), and provenance metadata for traceability. The deterministic identifiers support cross-referencing. To ensure structural integrity, the outputs are validated and corrected using the \textit{json\_repair} library\footnote{github.com/mangiucugna/json\_repair}.

\subsection{Multi-hop Inference for Synthetic Data Utility Evaluation}
% This is based on our previous paper https://arxiv.org/abs/2506.08713 should be an evaluation methods on the interpretable evaluation methods

Following the NLI-based assurance reasoning framework proposed in our previous work \cite{ikhwantri2025explainable}, we introduce a practical \textit{utility evaluation scenario} designed for the industrial problem, where real-world training data is unavailable. This methodology tests whether a model trained on synthetic data from Agentic RAG is acceptable by an external assessment body. For each CAE edge, the supporting child is the premise and its parent the hypothesis (e.g., Evidence $E \rightarrow$ Claim $C$), forming an Entailment pair. Not-Entailment pairs sample nodes without direct support, e.g., $E \rightarrow$ an unrelated claim $C'$ at the same hierarchical level.

We employ synthetic data generation for utility learning and evaluation~\cite{ji2025evaluating} to simulate the deployment of the framework in an SME context. This approach \textbf{(Synthetic-to-Real)} consists of two phases:

\noindent\textbf{Phase 1: Synthetic Training ($\mathcal{D}_{syn}$).} 
We utilise the Agentic RAG framework to process internal technical documentation for the PATROLIoT system, generating tree-based CAE structures grounded in local evidence. To construct the training corpus, we traverse the hierarchical edges of the generated CAE trees to produce NLI pairs. The textual content for each premise and hypothesis is extracted directly from the \textit{description} and \textit{rationale} fields defined in our argumentation schema (Section~\ref{sec: output-schema}), effectively transforming the structured argumentation logic into a dataset of natural language inference steps.

\noindent\textbf{Phase 2: Utility Testing ($\mathcal{D}_{real}$).} 
To validate the framework's practical utility, we leverage assessment records from an independent auditor, representing an external perspective. This dataset consists of real-world technical findings formatted as structured tabular data, comprising \textit{Auditor Comments}, \textit{SME Responses}, and \textit{Identified Gaps from Evidence}. We transform these records into a multi-hop natural language inference task designed to measure the model's ability to verify the consistency of the industrial audit trail. This evaluation phase determines how well the logic learned from synthetic training generalises to the rigorous requirements of an independent assessment.

\noindent Both of these phases, $\mathcal{D}_{syn}$ and $\mathcal{D}_{real}$, concern PATROLIoT.

\noindent\textbf{Inference variants.} \textbf{Chain} provides the complete reasoning path, $\text{[Premise]} + \text{[Intermediate Premises]} \rightarrow \text{[Hypothesis]}$, whereas \textbf{w/o Chain} provides only $\text{[Premise]} \rightarrow \text{[Hypothesis]}$, requiring intermediate steps to be inferred implicitly.

% \paragraph*{\textbf{Multi-hop Tabular Inference Formulation}}
% In the utility testing, we focus on the sequential reasoning chains found within industrial audit tables. We transform the tabular data, comprising \textit{Auditor Comments, SME Responses, and Identified Gaps from Evidence}, into a multi-hop natural language inference task. This measures the model's ability to verify the consistency of the audit trail.

% \textbf{We evaluate the utility of the NLI model by decomposing the existing requirement column into these discrete logical hops}. This provides automated sanity checks for complex compliance documentation similar to an assurance case.

\subsection{Expert Plausibility Study}
To evaluate the interpretability of our NLI evaluator, we conducted an expert plausibility study. The goal was not to validate the technical security of the product, but to assess whether the NLI model’s internal reasoning, represented via token-level heatmaps, aligns with the domain expert. 

The PATROLIoT Product Owner (PO) performed an audit of 60 NLI instances using an \textit{Offline Heatmap Evaluation} interface. For each sample, the label is provided (Entailment/Not-Entailment) and an explanation (highlighted tokens). The PO evaluated the plausibility of the explanation consisting of four questions with a 5-point Likert scale, and one multiple-choice question based on the following rubric:

\begin{description}
    \item[\textbf{Q1:}] (Justification) Do the highlighted tokens represent the technical keywords that actually justify the relationship? 
    \item[\textbf{Q2:}] (Necessity) Would the prediction be less reliable if these specific technical terms were ignored?
    \item[\textbf{Q3:}] (Sufficiency) Are the highlights enough to explain the logic without reading the entire document?
    \item[\textbf{Q4:}] (Trust) Does the quality of the highlighting increase trust in the NLI evaluator's decision?
    \item[\textbf{Q5:}] (Agreement) Do you think the model prediction is technically correct? (\textit{Yes / No / Unsure}). This question measures the alignment between the NLI model's classification and the Product Owner's ground-truth judgment. 
\end{description}

\section{Experimental Settings}
\subsection{Research Questions}

\begin{description}
    \item[\textbf{RQ1}] (Grounding and Traceability): How effectively does the Agentic RAG framework navigate heterogeneous documentation to establish verifiable, grounded links between generated assurance claims and technical evidence?
    % hallucination, citation coverage
    \item[\textbf{RQ2}:] (Inference Generalisation) How accurately does an NLI model trained on synthetic, agentically-generated documentation ($D_{syn}$) generalise to unseen, structured industrial assessment artefacts ($D_{real}$)?
    \item[\textbf{RQ3}:] (Explanation Plausibility): To what degree do the NLI model's rationales (token-level highlights) align with the technical knowledge of a domain expert?
\end{description}

\subsection{Models Configuration}
To evaluate the contribution of agentic grounding to the NLI evaluator's performance, we implement a baseline training scenario on different synthetic training data:

\begin{description}
    \item[\textbf{Vanilla-LLM ($D_{vanilla}$):}] We prompt a standard LLM to generate synthetic NLI pairs for CRA compliance without access to the PATROLIoT technical documentation. The model relies entirely on its internal parametric knowledge to synthesise technical claims and evidence.
    \item[\textbf{Agentic RAG (Proposed):}] The NLI evaluator is trained on $D_{syn}$, which is grounded in specific, retrieved technical artefacts via our agentic framework.
\end{description}

For fair comparison, we used \textbf{Qwen3-Coder-30B-A3B-Instruct-FP8}\footnote{https://huggingface.co/Qwen/Qwen3-Coder-30B-A3B-Instruct-FP8} for both Vanilla-LLM and Agentic-RAG models.

For the \textbf{NLI evaluation}, we compared two architectures: \textbf{BERT}\footnote{https://huggingface.co/google-bert/bert-large-uncased} for supervised fine-tuning and \textbf{LLaMA}\footnote{https://huggingface.co/meta-llama/Llama-3.2-1B} for In-Context Learning (ICL). The 1B Llama model was specifically chosen to ensure that gradient-based explanation extraction, such as Integrated Gradients, remains computationally feasible, allowing us to audit the model's focus on technical keywords. For the \textbf{expert plausibility study}, we use \textbf{Qwen2.5-32B-Instruct}\footnote{https://huggingface.co/Qwen/Qwen2.5-32B-Instruct} as the judge LLM. 

Both the Agentic RAG pipeline and the NLI models were executed
locally, ensuring that proprietary PATROLIoT documentation was
not transmitted to external model services.

\subsection{Requirement Split for Dataset Configuration.}
To evaluate the framework's generalisation capability, we implemented a requirement-wise data split across seven selected CRA requirements:
% \begin{description}
    % \item[\textbf{Training Set ($D_{syn}$ and $D_{real}$):}] 
    
    \textbf{Training Set ($D_{syn}$ and $D_{real}$):} Four requirements from the CRA were used to generate a synthetic corpus via the Agentic RAG output as synthetic Training. We select the four requirements from the cybersecurity requirements in Annex I part II of the EU CRA. In total, from 40 assurance cases, we convert into $\approx$4.2k instance pairs for the \textbf{Chain} setting, and $\approx$2.7k instance pairs for the \textbf{w/o Chain} setting.
    
    % \item[\textbf{Test Set ($D_{real}$):}] 
    \textbf{Test Set ($D_{real}$):} The model was evaluated on the remaining three unseen requirements from real-world tabular data from an independent audit as the \textbf{Utility Testing}\label{term:util-test}. This set comprises two cybersecurity requirements (Annex I part I) and one requirement about vulnerability management (Annex I part II). In total, from 3 requirements from tabular data, we convert them into 30 instance pairs for the Chain setting with at most 4 hops and 24 instance pairs for the w/o Chain setting.
% \end{description}

% The test set serves as a form of unseen requirements setup to evaluate the Multi-hop Inference model's capacity for cross-requirement generalisation. 
This requirement-wise split evaluates generalisation to regulatory requirements unseen during training. It simulates a realistic industrial scenario where an SME must apply learned compliance logic to new or evolving regulatory articles without access to pre-labelled training data for those specific mandates.

\section{Results and Discussion}

\subsection{Grounding and Traceability: Citation Coverage (\textbf{RQ1})}

We define \textbf{citation coverage} as the ratio of substantiated evidence nodes, those containing direct provenance links to technical documentation, relative to the total number of evidence nodes generated. This metric measures traceability, not evidence correctness or overall assurance-case validity.

% Across 70 generated ACs, the Agentic RAG produced 356 evidence nodes with 307  substantiated artefact references, resulting in \textbf{citation coverage of $\approx$4.4 artefacts mentioned per AC}. 
Across 70 generated ACs, 307 of 356 evidence nodes contained artefact references, corresponding to 86.2\% citation coverage and $\approx$4.4 referenced artefacts per AC. The discrepancy (13.8\%, i.e., 49/356) where evidence nodes lacked specific artefact links serves as a valuable diagnostic signal for documentation gaps, proactively identifying requirements that lack technical substantiation. These 307 references originated from 42 unique source entries, only \textbf{22 of which (52.3\%, i.e., 22/42) contained formal file extensions}. This variability underscores the agentic framework's ability to maintain high traceability while navigating the heterogeneous and inconsistent metadata of industrial SME repositories.

\subsection{Inference Generalisation (\textbf{RQ2})}

To evaluate the effectiveness of the NLI evaluator, we report performance across two primary training scenarios. The first scenario utilises real-world data for both training and testing to establish a performance ceiling. The second scenario evaluates the practical utility of our synthetic training strategy when applied to unseen real-world assessment artefacts.

\noindent\textbf{Baseline: Train (Real) $\rightarrow$ Test (Real).}
This represents the traditional supervised approach using the limited human-labelled audit data available. As shown in Table~\ref{tab:merged_nli_results}, this configuration suffers from extreme data scarcity, leading to poor generalisation. In many cases, the model fails to learn the underlying compliance logic, instead over-fitting to the specific linguistic noise of the small sample.

\begin{table}[ht]
\centering
% \small
\footnotesize
\caption{NLI Performance comparison across Training Sources and Reasoning Settings (Chain vs. w/o Chain).}
\label{tab:merged_nli_results}
\begin{tabular}{@{}l l l c c@{}}
\hline
\textbf{Model} & \textbf{Training Source} & \textbf{Setting} & \textbf{Acc.} & \textbf{F1} \\ \hline
\multirow{5}{*}{BERT}  & Real (Baseline)    & Chain     & 0.53 & 0.23 \\
                       & Real (Baseline)    & w/o Chain & 0.54 & 0.23 \\ \cline{2-5} 
                       & Vanilla LLM        & Chain     & 0.70 & 0.69 \\ 
                       % \cline{2-5} 
                       & \textbf{Agentic RAG} & Chain     & 0.83 & 0.82 \\
                       & \textbf{Agentic RAG} & \textbf{w/o Chain} & \textbf{0.88} & \textbf{0.85} \\ \hline
\multirow{5}{*}{LLaMA} & Real (Baseline)    & Chain     & 0.37 & 0.31 \\
                       & Real (Baseline)    & w/o Chain & 0.50 & 0.47 \\ \cline{2-5} 
                       & Vanilla LLM        & Chain     & 0.43 & 0.31 \\ 
                       % \cline{2-5} 
                       & \textbf{Agentic RAG} & Chain     & 0.53 & 0.37 \\
                       & \textbf{Agentic RAG} & w/o Chain & 0.54 & 0.38 \\ \hline
\end{tabular}
\end{table}

\noindent\textbf{Utility Testing: Train (Synthetic) $\rightarrow$ Test (Real).}
This scenario represents the core industrial challenge where real-world training data is unavailable. We evaluate the utility of synthetic data generated via a \textit{Vanilla LLM} versus our \textit{Agentic RAG} framework.

To measure the practical utility of the NLI evaluator, we report its performance on the unseen auditor dataset ($D_{real}$) after training on the synthetic SME-side documentation ($D_{syn}$). 

\noindent\textbf{Classification Performance:} The model achieved an accuracy of $88\%$ and an F1-score of $85\%$. These results indicate that the "internal" logic learned from the Agentic RAG-generated samples translates effectively to the "external" perspective of independent audit tables. The results in Table~\ref{tab:merged_nli_results} provide several key insights. 
 
 \textbf{Impact of Grounding:} Models trained on Agentic RAG data outperformed the ungrounded Vanilla-LLM baseline, indicating a benefit from document grounding. This comparison does not isolate agentic iteration from conventional RAG.
 
 % This confirms that grounding synthetic generation in local technical documentation is essential for the NLI model to learn the specific semantic relationships of the industrial product rather than relying on hallucinated generic patterns.

\textbf{Value of Chain Reasoning:} The NLI model with the \textbf{Chain} setting generally yielded lower performance than the model in the \textbf{w/o Chain} setting (e.g., 0.83 vs 0.88 for BERT). This shows that the synthetic training data of hierarchical Claim-Argument-Evidence trees can be linearised into a multi-hop representation, which is also applicable to real-world multi-hop tabular data. The models trained in direct settings generalise more effectively to these tabular audit records, whereas multi-hop logic may introduce cumulative error or over-specialisation to hierarchical structures.

\textbf{Fine-tuned BERT vs ICL LLaMA :} Fine-tuned BERT outperformed the 1B LLaMA ICL configuration by up to 35\%. Since the models differ in both adaptation strategy and capacity, these results do not isolate an encoder-versus-decoder effect.

% The BERT model showed a significant advantage in discriminative accuracy and F1-score, outperforming LLaMA by nearly 35\% in the best case. In contrast, Llama-1B were underpowered for this specific discriminative task, highlighting the trade-offs between specialised NLI fine-tuning BERT and the pretrained decoder-only models LLaMA.

\subsection{Explanation Plausibility (\textbf{RQ3})}

\begin{table}[ht]
\centering
% \footnotesize
\scriptsize
\caption{Mean plausibility scores (Q1--Q4, 1--5) and Q5 category proportions (\%). 
% Skipped instances are defined from human \texttt{q6\_comment} and removed from both human and LLM rows.
}
\label{tab:plausibility_comparison_human_llm}
\begin{tabular}{@{}llcccc@{}}
\toprule
\multirow{2}{*}{\textbf{Source}} & \multirow{2}{*}{\textbf{Metric}} & \multicolumn{2}{c}{\textbf{BERT}} & \multicolumn{2}{c}{\textbf{LLaMA}} \\
% \cmidrule(lr){3-4} \cmidrule(lr){5-6}
& & \textbf{Chain} & \textbf{w/o Chain} & \textbf{Chain} & \textbf{w/o Chain} \\
\midrule
% Human & N (kept) & 18 & 12 & 18 & 12 \\
% \midrule
Human & Q1: Just. & 3.06 & 3.25 & 2.17 & 1.83 \\
Human & Q2: Nec. & 2.78 & 3.17 & 2.94 & 2.58 \\
Human & Q3: Suf. & 2.94 & 3.33 & 1.78 & 1.75 \\
Human & Q4: Trust & 2.78 & 2.50 & 1.78 & 1.58 \\
Human & \textbf{Mean} & \textbf{2.89} & \textbf{3.06} & \textbf{2.17} & \textbf{1.94} \\
Human & Q5 Agree. (\%) & 50.0 & 41.7 & 16.7 & 25.0 \\
% Human & Q5 N used & 18 & 12 & 18 & 12 \\
% LLM Judge & N (kept) & 18 & 12 & 18 & 12 \\
\midrule
% LLM Judge & N (kept) & 18 & 12 & 18 & 12 \\
% \midrule
LLM Judge & Q1: Just. & 3.11 & 3.17 & 3.00 & 3.25 \\
LLM Judge & Q2: Nec. & 3.39 & 3.92 & 3.89 & 3.50 \\
LLM Judge & Q3: Suf. & 3.00 & 3.00 & 2.78 & 3.08 \\
LLM Judge & Q4: Trust & 3.28 & 3.42 & 3.00 & 3.25 \\
LLM Judge & \textbf{Mean} & \textbf{3.19} & \textbf{3.38} & \textbf{3.17} & \textbf{3.27} \\
LLM Judge & Q5 Agree. (\%) & 11.1 & 25.0 & 38.9 & 33.3 \\
% LLM Judge & Q5 N used & 18 & 12 & 18 & 12 \\
\bottomrule
\end{tabular}
\end{table}

Our expert plausibility study reveals a significant interpretability gap between the BERT model with fine-tuning and the LLaMA with the ICL approach, with human auditors favouring BERT’s technical faithfulness. Human experts awarded BERT a peak mean plausibility score of 3.06 and a 50\% technical agreement rate (Q5), whereas LLaMA’s explanations were deemed insufficient and poorly justified, scoring as low as 1.75 in sufficiency. Conversely, the LLM Judge exhibited a notable "plausibility bias," consistently overrating LLaMA’s justifications and failing to align with human experts on technical correctness, which is shown by the sharp discrepancy in BERT Chain agreement (11.1\% LLM vs 50\% Human). These results demonstrate that while ICL-based explanations may appear linguistically plausible to automated judges, supervised fine-tuning on agentically grounded data can provide more technically aligned token-level justifications in this setting.

% Report the mean scores from the User Evaluation Study (Q1–Q4).

% Rating Summary (Scaled 1–5):
% Example
%     Q1 (Justification): 3.8 — Indicates the highlights generally align with expert reasoning.

%     Q2 (Necessity): 3.2 — Suggests moderate organisational necessity of highlighted tokens.

%     Q3 (Sufficiency): 3.5 — Reflects that highlights provide a reasonable basis for the prediction.

%     Q4 (Trust): 3.7 — Shows the interface improves expert confidence in the automated assessment.

\subsection{Discussion}

\textbf{Implications for Practitioners.} Our agentic RAG addresses the labour-intensive self-assessment bottleneck for SMEs in CRA by automating documentation retrieval across fragmented documents, then converting the CRA requirements into assurance cases. In practice, our citation coverage metric could be used to identify documentation gaps before formal audits to reduce the risk of certification failure. Our NLI evaluation tools could enable semi-automatic workflows with the Agentic RAG, such as pruning the redundant or irrelevant nodes generated by the LLMs. Our evaluation with domain experts shows that the proposed approach provides traceability from input tokens to generated outputs, supporting transparency required for regulatory compliance in the European market.

\textbf{Iterative Refinement and Security Level.} 
The industrial case study highlights that generating ACs is rarely a one-shot process. Initial outputs often reveal a need to refine the arguments or provide more granular information regarding the product's security level. Although our framework fulfils this requirement through its agentic, iterative capabilities, achieving compliance with higher security levels requires updating the evidence based on real-world practice or implementation rather than further assurance case refinement. 

\subsection{Threats to Validity}

\textbf{Construct Validity.} This study distinguishes between technical logical soundness and formal regulatory compliance. While domain experts validated the semantic consistency of the generated arguments, they are not legal authorities on the CRA. Consequently, while the system produces logically coherent evidence, it cannot guarantee legal audit success; future work must involve legal practitioners to ensure formal compliance-readiness.

\textbf{Internal Validity.} Subjective bias poses a primary threat to internal validity since the domain expert evaluator also co-authors this paper. This dual role creates a risk of confirmation bias where the evaluator might inadvertently favour model outputs. To mitigate this, we cross-referenced expert ratings with an independent LLM-as-a-Judge uninvolved in the generation process. Reporting the agreement between the human expert and the automated judge ensures a more objective assessment of system performance.

\textbf{External Validity.} External validity is limited by the single-product focus of this study. While providing deep insight into an SME's needs, the findings may not generalise across all industrial sectors. We addressed this by focusing on the logic of NLI 'plausibility' (Q1-Q4) rather than an absolute ground truth of compliance, which remains a matter of legal interpretation.

\section{Conclusion}

% This paper introduced an Agentic RAG framework for automated Assurance Case generation and evaluation under the EU CRA. Our results demonstrate that the framework effectively addresses the industry problem, where real-world training data is not always available at the beginning. To address \textbf{RQ1}, our Agentic RAG automates the slow process of manually going through heterogeneous documentation, achieving a high grounding density ($\approx$4.4 artefacts per AC) and replacing manual search with a traceable audit trail. Findings for \textbf{RQ2} show that agentically grounded synthetic data allows models to learn regulatory logic without requiring expertise-dependent and manually labelled datasets. Evaluation of \textbf{RQ3} confirms that fine-tuned encoder models provide the technical justifications necessary for engineers to analyse compliance states efficiently. Our approach streamlines the certification process and removes the significant barriers to timely industrial compliance by shifting the burden from manual efforts to automated and interpretable reasoning.

This paper introduced an Agentic RAG framework for automated Assurance Case generation and evaluation under the EU CRA. Our results demonstrate that the framework effectively addresses the industry problem, where real-world training data is not always available at the beginning. \textbf{RQ1:} Agentic RAG automates document retrieval and AC construction while preserving provenance traceability across heterogeneous technical documentation. \textbf{RQ2:} Agentically grounded synthetic data supports generalisation to unseen auditor requirements when labelled real-world data are scarce. \textbf{RQ3:} Interpretable NLI can support engineers in inspecting the technical basis of compliance-related decisions. Overall, the framework provides traceable and interpretable support for CRA assurance workflows, while formal compliance remains subject to expert assessment.

\paragraph*{Code and Data Availability}
The NLI training code and the evaluation interface are publicly accessible (available via\footnote{https://github.com/farizikhwantri/exclaim}~\cite{ikhwantri2025explainable}). Due to confidentiality with our industry partner, the Agentic RAG implementation, the output and technical datasets remain proprietary.

\section*{Acknowledgment}

This work is funded by the European Commission under grant agreement No. 101120606, CertifAI. We thank EZU for the evaluation of the Catalink assessment data. This work is supported by the eX3 Infrastructure, funded by the Research Council of Norway under contract 270053.

% \bibliographystyle{IEEEtran}
% \bibliography{references}

\end{document}